\documentstyle[12pt]{article}
\begin{document}
\thispagestyle{empty}
\begin{center}
\LARGE \tt \bf {Gravitational and torsion waves in linearised teleparallel gravity}
\end{center}

\vspace{3.5cm}

\begin{center}
{\large By L.C. Garcia de Andrade\footnote{Departamento de F\'{\i}sica Te\'{o}rica - IF - UERJ - Rua S\~{a}o Francisco Xavier 524, Rio de Janeiro, RJ, Maracan\~{a}, CEP:20550.e-mail:garcia@dft.if.uerj.br}}
\end{center}

\begin{abstract}
Spin-2, spin-1 and spin-0 modes in linearised teleparallelism are obtained where the totally skew-symmetric complex contortion tensor generates scalar torsion waves and the symmetric contortion in the last two indices generates gravitational waves as gravitational perturbations of flat spacetime with contortion tensor. A gedanken experiment with this gravitational-torsion wave hitting a ring of spinless particles is proposed which allows us to estimate the contortion of the Earth by making use of data from LISA GW detector. This value coincides with previous value obtained by Nitsch in teleparallelism using another type of experiment. 

\end{abstract}

\newpage

\section{Introduction}
Since the early proposals of experiments \cite{1,2,3,4,5} to detect spacetime torsion or Cartan contortion \cite{6} and to give an experimental basis to alternative gravity theories of Einstein-Cartan type \cite{7} theoretical physicists working on the field have noticed that torsion detection by the technique of spin precession have been proved hopeless since all computations are very far beyond all quantum capabilities of laboratory devices and even astrophysical probes would be difficult to construct for this task. In this letter we propose a new method to estimate spacetime contortion which is based on the GW detectors in linearised teleparallel space. This space which has been applied to gravity by Einstein in 1932 in Berlin \cite{8} in the attempt to build a unified geometrical theory of gravitation and electrodynamics proved recently \cite{9} to be very important in solving theoretical problems in gravitation such as the tetrad complex of gravitational energy. This space is used here with the aim of simplifying the computations since by definition teleparallel spacetime possess a total Riemann-Cartan curvature that vanishes in all spacetime manifold. Besides the use of teleparallelism allows us to obtain a contortion wave based on the axion contortion which is by definition a totally skew-symmetric Cartan contortion tensor. This axion contortion in linearized gravity generates only scalar waves for the torsion potential and the symmetry of contortion in the last two indices generates  gravitational waves as perturbations of Minkowski flat which we call by the collective name of linearised gravitational-torsion waves. A gedanken experiment is obtained by sending this gravitational-contortion waves to a ring of spinless particles and showing that the contortion waves act as a damping force on the spinless particles. The idea of interpreting the contortion as a force on the RHS of the nongeodesic equation in teleparallel gravity has been presented recently by de Andrade et al. \cite{10}. The connection between spinless particles and torsion or contortion has also been provided recently in $T_{4}$ by de Andrade et al. \cite{10} and by Kleinert \cite{11} in the context of Riemann-Cartan geometry.
The paper is organized as follows. In the section 2 we present the idea and mathematical computations releted to the gravitational-torsion waves in $T_{4}$ spacetime. In section 3 we propose that Laser interferometer GW observatories (LISA) could be used along with the Nitsch result of contortion at the surface of the Earth, which is of the  order of $10^{-24} s^{-1}$ to obtain a metric perturbation amplitudes from this data and compare it with the GW detector result. In the last section we discuss future prospects. It is important to stress yet that here we are not considering gravitational or torsion waves in metric affine theories of gravity as  treated recently in the literature \cite{12,13}. We also pointed out that since teleparallelism is equivalent to general relativity (GR), strictly speaking we do not detect torsion but this yields a powerful method to estimate the value for torsion in various situations linked with GW detectors. Besides as far as we know Smalley \cite{14} has shown that the experimental equivalence between $T_{4}$ theory of gravity and GR is valid up to the order $O(v^{5})$ in the velocities.
\section{Gravitational-torsion waves in teleparallel spacetime}
Before we start we should like to mention that although neutrinos are not compatible with $T_{4}$ geometry as shown by Letelier \cite{15} the neutrinos maybe consider as a torsion wave but not all torsion waves are neutrino waves. Actually neutrino geometry with torsion considered earlier by Kuchowicz \cite{16} has been obtained exactly considering a totally skew-symmetric contortion exactly as here. But of course neutrino waves would obey also the Weyl neutrino equation besides the Einstein-Cartan equations or the teleparallel constraint considered here. We also note that the idea of obtained massless gravitons from linearized teleparallelism was previously obtained by M\"{u}ller-Hoissen and Nitsch \cite{17} where pp gravitational wave solutions were extended to teleparallel gravity. Our approach here is slightly distinct since we consider complex contortion which are also axionic which allows us to obtain more degrees of freedom to obtain scalar torsion waves as well and also to get a touch of GW experiments data to estimete contortion tensor. Let us consider the RC curvature tensor in the linearised form \cite{11}
\begin{equation}
R_{ijkl}= {R^{0}}_{ijkl}+{\partial}_{i}A_{jkl}- {\partial}_{j}A_{ikl}
\label{1}
\end{equation}
where the upper indice $(0)$ denotes the Riemannian objects while the mathematical objects without indices denote the RC objects. Thus by making use of the teleparallel constraint
\begin{equation}
{R_{ijkl}}({\Gamma})= 0
\label{2}
\end{equation}  
where ${\Gamma}$ is the RC connection of spacetime manifold. By substituting this value into expression (\ref{1}) one obtains
\begin{equation}
{R^{0}}_{ijkl}= -{\partial}_{i}A_{[jkl]}+ {\partial}_{j}A_{[ikl]}
\label{3}
\end{equation}
By considering the totally skew-symmetric part of Cartan contortion 
\begin{equation}
A_{[ijl]}= {\epsilon}_{ijkl}{\partial}^{l}{\phi}
\label{4}
\end{equation}
where ${\epsilon}_{ijkl}$ is the Levi-Civita totally skew-symmetric symbol and ${\phi}$ is the torsion potential of the axion contortion (\ref{4}). Substitution of expression (\ref{4}) into (\ref{3}) yilds
\begin{equation}
{R^{0}}_{ijkl}= -{\partial}_{i}{\epsilon}_{jklp}{\partial}^{p}{\phi}+ {\partial}_{j}{\epsilon}_{iklp}{\partial}^{p}{\phi}
\label{5}
\end{equation}
since the Riemann curvature tensor is symmetric in the first pair of indices , contraction expression (\ref{5}) with the appropriated  Levi-Civita symbol one obtains that   
\begin{equation}
{R^{0}}_{ijkl}{\epsilon}^{ijkl}=0
\label{6}
\end{equation}
Substitution of this expression into the  contacted (\ref{5}) one obtains the linearised torsion wave equation ${\Box}{\phi}=0$ ,(where ${\Box}={\partial}^{i}{\partial}_{i}$ is the Minkowskian D'Lambertian), for the scalar potential which represents the spin-0 mode for contortion. Let us consider now the contortion components which are symmetric in the last two indices given by $K_{i(jk)}$ and built the gravitational wave tensor perturbations given by
\begin{equation}
g_{ij}= {\eta}_{ij}+ h_{ij}
\label{7}
\end{equation}
where ${\eta}_{ij}$ is the Minkowski metric and $h_{ij}$ where $|h_{ij}|<<1$ are the gravitational perturbations. By considering now the first-order Ricci components in $U_{4}$ RC spacetime
\begin{equation}
{{R^{(1)}}_{ij}}(\Gamma)= {{R^{0}}^{(1)}}_{ij}-{\partial}_{j}{K^{l}}_{il}+ {\partial}_{l}{K^{l}}_{ij}
\label{8}
\end{equation}
Note that in teleparallelism the vanishing of the RC curvature implies the vanishing of Ricci-Cartan curvature (\ref{8}) which reduces this expression to 
\begin{equation}
{{R^{0}}^{(1)}}_{ij}=+{\partial}_{j}{K^{l}}_{il}- {\partial}_{l}{K^{l}}_{ij}
\label{9}
\end{equation}
by making the choice of the gauge contortion as ${K^{l}}_{il}=0$ which says that the totally-skew symmetric part of the contortion vanishes, one obtains from (\ref{9}) yields
\begin{equation}
{{R^{0}}^{(1)}}_{ij}= - {\partial}_{l}{K^{l}}_{ij}
\label{10}
\end{equation}
By using the appropriate Lorentz gauge in linearised GR the Riemannian part of the first-order perturbation Ricci tensor is 
\begin{equation}
{{R^{0}}^{(1)}}_{ij}= - \frac{1}{2}{\Box}h_{ij}
\label{11}
\end{equation}
Substitution of this expression into formula (\ref{10}) yields
\begin{equation}
{\Box}h_{ij}= 2{\partial}_{l}{K^{l}}_{ij}
\label{12}
\end{equation}
This equation shows that GW in linearised teleparallelism have the divergence of the non-totally skew-symmetric contortion $K_{ijl}$ as a source for the GW. To end this section we shall show that the scalar torsion wave equation maybe obtained also in the teleparallel version of other theories of propagating torsion as in the theory of Hojmann, Rosenbaum and Ryan (HRR) \cite{18} on their investigation of propagating torsion theory with a scalar field mechanism which works as a torsion potential and provides the mechanism to generate the torsion waves in RC spacetime. Before address our main problem here we show that the teleparallel constraint applied to HRS theory yields a scalar torsion wave on the spacetime background. To this end let us reproduce here the expression for the RC curvature tensor found by HRR \cite{18} in the flat space approximation where $g_{ij}={\eta}_{ij}$
\begin{equation}
R^{i}_{jkl}({\Gamma}')= \frac{\sqrt{G}}{\sqrt{3}c^{2}}[{\eta}_{jl}{\partial}^{i}{\partial}_{k}{\phi}-{\delta}^{i}_{l}{\partial}^{j}{\partial}_{k}{\phi} + {\delta}^{i}_{k}{\partial}_{j}{\partial}_{l}{\phi} -{\eta}_{jk}{\partial}^{i}{\partial}_{l}{\phi}]
\label{13}
\end{equation}
Double appropriate contraction of this last expression yields the Ricci-Cartan scalar
\begin{equation}
R({\Gamma}')=6\frac{\sqrt{G}}{\sqrt{3}c^{2}}\Box{\phi}
\label{14}
\end{equation}
Once again we make expression (\ref{14}) to vanish to be compatible with the $T_{4}$ condition. This reduces this expression to
\begin{equation}
\Box{\phi}=0
\label{15}
\end{equation}
represents the plane scalar torsion wave in teleparallel background. Let us now consider the metric perturbation or GW geometry \cite{19} as a perturbation of flat spacetime metric ${\eta}_{ij}$  as before. In this case the HRR theory of propagating torsion reduces to the Riemann curvature in terms of the scalar torsion potential ${\phi}$
\begin{equation}
R^{i}_{jkl}({\Gamma})=-\frac{\sqrt{G}}{\sqrt{3}c^{2}}[{\eta}_{jl}{\partial}^{i}{\partial}_{k}{\phi}+{\delta}^{i}_{l}{\partial}_{j}{\partial}_{k}{\phi} -{\delta}^{i}_{k}{\partial}^{j}{\partial}_{l}{\phi} +{\eta}_{jk}{\partial}^{i}{\partial}_{l}{\phi}]
\label{16}
\end{equation}
As we shall see bellow in a simpler case by applying the weak field approximation here one would obtain an expression between the small amplitude perturbation and the scalar torsion potential ${\phi}$. Let us now consider $T_{4}$ simplest model where no scalar field is present which is enough for our purposes to estimate the amplitude of perturbations of spacetime in terms of contortion. This model yields the following Riemann curvature tensor w.r.t contortion
\begin{equation}
R^{a}_{{0}{b}{0}}(g)={\partial}^{a}{K_{0b 0}}-{\partial}_{0}{{K}^{a}}_{b 0}
\label{17}
\end{equation}
Bwhere here g is the symbolic representation of the metric tensor $g_{ij}$. By considering that the contortion is just a function of time and that the metric curvature can be expressed as 
\begin{equation}
R_{a0b0}(g)=-\frac{1}{2}{{\ddot{h}}^{TT}}_{ab}= -{\partial}_{0}{K_{ab0}}
\label{18}
\end{equation}
which allows us to write the contortion in terms of the acceleration of metric perturbation h as
\begin{equation}
\frac{1}{2}{{\dot{h}}^{TT}}_{ab}= {K_{ab0}}
\label{19}
\end{equation}
which allows us to find the tensor or GW perturbation in terms of the Cartan contortion tensor as
\begin{equation}
{{h}^{TT}}_{ab}= 2\int{{K_{ab0}}dt}
\label{20}
\end{equation}
In the next section we shall consider the application of the theory discussed in this section.
\section{GW and the motion of spinless particles in teleparallelism}
Let us consider the propagating along the $z-axis$, the proper distance in the $xy$-plane \cite{18} is given by
\begin{equation}
dl = [(1+{h^{TT}}_{xx})dx^{2}+ (1-{h^{TT}}_{xx})dy^{2}+2 {h^{TT}}_{xy}dx dy]^{\frac{1}{2}}
\label{21}
\end{equation}
where the metric perturbation components are
\begin{equation}
{h^{TT}}_{xx}= -{h^{TT}}_{yy}=A_{+} e^{-i{\omega}(t-z)}
\label{22}
\end{equation}
\begin{equation}
{h^{TT}}_{xy}= A_{X} e^{-i{\omega}(t-z)}
\label{23}
\end{equation}
The Riemann curvature components are obtained by making use of the $T_{4}$ condition on vanishing of the RC curvature tensor. This yields 
\begin{equation}
{R^{0}}_{1010}=[{\partial}_{1}K_{010} - {\partial}_{0}K_{110}]
\label{24}
\end{equation}
\begin{equation}
{R^{0}}_{1310}=[{\partial}_{3}K_{110} - {\partial}_{1}K_{310}]= -\frac{1}{2}\ddot{h_{+}}= -\frac{{\omega}^{2}}{2}A_{+}e^{-i{\omega}(t-z)}
\label{25}
\end{equation}
\begin{equation}
{R^{0}}_{1313}=[{\partial}_{3}K_{311} - {\partial}_{1}K_{313}] 
\label{26}
\end{equation}
\begin{equation}
{R^{0}}_{2020}=-[{\partial}_{0}K_{220} - {\partial}_{2}K_{020}]
\label{27}
\end{equation}
\begin{equation}
{R^{0}}_{2320}=[{\partial}_{3}K_{220} - {\partial}_{2}K_{320}]
\label{28}
\end{equation}
\begin{equation}
{R^{0}}_{2323}=[{\partial}_{3}K_{223} - {\partial}_{2}K_{323}]
\label{29}
\end{equation}
\begin{equation}
{R^{0}}_{1020}=[{\partial}_{0}K_{120} - {\partial}_{1}K_{020}]= -\frac{1}{2}{\ddot{h}}_{X}
\label{30}
\end{equation}
\begin{equation}
{R^{0}}_{1320}=[{\partial}_{3}K_{120} - {\partial}_{1}K_{320}] 
\label{31}
\end{equation}
\begin{equation}
{R^{0}}_{1023}=[{\partial}_{3}K_{120} - {\partial}_{1}K_{023}]= -\frac{1}{2}{\ddot{h}}_{X}
\label{32}
\end{equation}
\begin{equation}
{R^{0}}_{1323}=[{\partial}_{3}K_{123} - {\partial}_{1}K_{323}]= -\frac{1}{2}{\ddot{h}}_{X}
\label{33}
\end{equation}
Analogous expressions for the $h_{+}$ mode allows us to perform the following computations. Since the computations are analogous we do not make all of them here but is enough to show that it works in some cases. Thus let us consider the case of the component 
\begin{equation}
{R^{0}}_{2020}=-[{\partial}_{0}K_{220} - {\partial}_{2}K_{020}]
\label{34}
\end{equation}
Since by definition $h(t-z)$ for both modes and $(x^{0}=t,x^{1}=x,x^{2}=y,x^{3}=z)$ this expression reduces to
\begin{equation}
{R^{0}}_{2020}=-{\partial}_{0}K_{220}= -\frac{1}{2}\ddot{h_{+}}
\label{35}
\end{equation}
By analogy considering the component 
\begin{equation}
{R^{0}}_{2320}={\partial}_{3}K_{220}= -\frac{1}{2}\ddot{h_{+}}
\label{36}
\end{equation}
From expressions (\ref{32}) and (\ref{33}) we obtain 
\begin{equation}
({\partial}_{0}+{\partial}_{3})K_{220}= 0
\label{37}
\end{equation}
By operating with expression ${\partial}_{0}-{\partial}_{3}$ on the LHS of equation (\ref{34}) one obtains the expression for the wave equation for the contortion component $K_{220}$  
\begin{equation}
({{\partial}^{2}}_{0}-{{\partial}^{2}}_{3})K_{220}= 0
\label{38}
\end{equation}
which can be written as ${\Box}{K_{220}}=0$. This wave equation can be shown to be valid for all other contortion components to yield the following contortion wave equation 
\begin{equation}
{\Box}K_{ijk}= 0
\label{39}
\end{equation}
Note that analogous equation for the totally skew-symmetric part of contortion $A_{[ijk]}$ maybe obtained as well as 
\begin{equation}
{\Box}A_{ijk}= 0
\label{40}
\end{equation}
which gives us the equivalent spin-1 equation
\begin{equation}
{\Box}A_{i}= 0
\label{41}
\end{equation}
where $A_{i}= {\epsilon}_{ijkl}A^{jkl}$. This is analogous to the wave equation for the electromagnetic vector potential. Taking for example the component $K_{123}$ one obtains the following complex solution
\begin{equation}
K_{123}= -\frac{i{\omega}}{2}A_{X}e^{-i{\omega}(t-z)}
\label{42}
\end{equation}
Note that this expression shows us that the contortion wave has a phase difference of $\frac{\pi}{2}$ w.r.t the GW. When one takes the real part of the complex phase this expression reduces to
\begin{equation}
Im(K_{123})=- \frac{A_{X}}{2}{\omega}cos({\omega}(t-z))
\label{43}
\end{equation}
where the Im denotes the imaginary part of the complex representation of the torsion wave. When the torsion wave passes by a ring of spinning particles the perturbation of the ring in the $xy$-plane is
\begin{equation}
{\delta}l^{X} = -\frac{A_{X}}{2}l cos({\omega}(t-z))
\label{44}
\end{equation}
Thus contortion contributes to ${\delta}l$ which is the change of the separation of the spinning particles when the torsion wave hits the ring. Actually is better to consider the relative displacement of particles in the ring of particles which is given by $\frac{{\delta}l}{l}$ which by the expressions (\ref{43}) and (\ref{44}) yields
\begin{equation}
|Im(K_{123})|= {\omega}\frac{{\delta}l^{X}}{l}
\label{45}
\end{equation}
Note that by considering a frequency of the order of ${\omega}=10^{-3}Hz$ and the value obtained by Nitsch \cite{12} for the contortion at the surface of the Earth of $K_{123}=10^{-24} s^{-1}$ one obtains from expression (\ref{45}) a relative displacement for the ring of 
\begin{equation}
\frac{{\delta}l^{X}}{l}=10^{-21}
\label{46}
\end{equation}
\newpage
\section{Conclusions}
We show that linearised teleparallelism contains spin-0, spin-2 and spin-1 tensor modes from the simple RC constraint of $T_{4}$ gravity. This structure allows us to propose a gedanken experiment to estimate the values of contortion. Note that the presence of the spins $0$ (dilatons or mesons for example),$1$ (massless vector particles analogous to the electromagnetic fields)  and $2$ (massless gravitons). In near future a more detail investigation of the non-geodesic equation of motion of spinless pearticles may shed some light on the structure of teleparallelism as a candidate to substitute the Einstein-Cartan theory of gravity.
\section*{Acknowledgements}
 I would like to express my gratitude to Prof. I.D. Soares and M. Novello for pointing out a mistake in the first draft of this paper. Thanks are also due to Prof. J.G.Pereira,Dr. Cristine. N. Ferreira and Prof. Ilya Shapiro for helpful discussions on the subject of this paper. Financial support from CNPq. is grateful acknowledged.

\end{document}